\documentclass[11pt,a4paper,reqno]{article}
\usepackage{mysty}
\usepackage{fullpage}

\usepackage[thinlines]{easytable}

\SetLabelAlign{parright}{\parbox[t]{\labelwidth}{\raggedleft{#1}}}
\setlist[description]{style=multiline,topsep=4pt,align=parright}

\makeatletter
\let\reftagform@=\tagform@
\def\tagform@#1{\maketag@@@{(\ignorespaces\textcolor{black}{#1}\unskip\@@italiccorr)}}
\newcommand{\iref}[1]{\textup{\reftagform@{\tcr{\ref{#1}}}}}
\makeatother

\newenvironment{affiliations}{%
    \setcounter{enumi}{1}%
    \setlength{\parindent}{0in}%
    \slshape\sloppy%
    \begin{list}{\upshape$^{\arabic{enumi}}$}{%
        \usecounter{enumi}%
        \setlength{\leftmargin}{0in}%
        \setlength{\topsep}{0in}%
        \setlength{\labelsep}{0in}%
        \setlength{\labelwidth}{0in}%
        \setlength{\listparindent}{0in}%
        \setlength{\itemsep}{0ex}%
        \setlength{\parsep}{0in}%
        }
    }{\end{list}\par\vspace{12pt}}
    
\begin{document}

\title{Determination of Effective Synaptic Conductances Using Somatic Voltage Clamp}
\author{Songting Li$^{1}$, Nan Liu$^2$, Xiaohui Zhang$^{2,*}$, Douglas Zhou$^{3,*}$, \& David Cai$^{1,3,4,*}$}

\date{}
\maketitle
\begin{affiliations}
 \item Courant Institute of Mathematical Sciences and Center for Neural Science, New York University, New York, NY, United States of America
 \item State Key Laboratory of Cognitive Neuroscience and Learning, IDG/McGovern Institute for Brain Research, Beijing Normal University, Beijing, China
 \item School of Mathematical Sciences, MOE-LSC, and Institute of Natural Sciences, Shanghai Jiao Tong University, Shanghai, China
 \item NYUAD Institute, New York University Abu Dhabi, Abu Dhabi, United Arab Emirates\\
\end{affiliations}

\begin{abstract}
The interplay between excitatory and inhibitory neurons imparts rich functions of the brain. To understand the underlying synaptic mechanisms, a fundamental approach is to study the dynamics of excitatory and inhibitory conductances of each neuron. The traditional method of determining conductance employs the synaptic current-voltage (I-V) relation obtained via voltage clamp. Using theoretical analysis, electrophysiological experiments, and realistic simulations, here we demonstrate that the traditional method conceptually fails to measure the conductance due to the neglect of a nonlinear interaction between the clamp current and the synaptic current. Consequently, it incurs substantial measurement error, even giving rise to unphysically negative conductance as observed in experiments. To elucidate synaptic impact on neuronal information processing, we introduce the concept of effective conductance and propose a framework to determine it accurately. Our work suggests re-examination of previous studies involving conductance measurement and provides a reliable approach to assess synaptic influence on neuronal computation.
\end{abstract}

Neurons receive myriad excitatory (E) and inhibitory (I) synaptic inputs at dendrites. The spatiotemporal interaction between these E and I inputs are crucial for neuronal computation\cite{marino2005invariant,yizhar2011neocortical,deneve2016efficient}, for instance, to shape neural activity \cite{wehr2003balanced,zerlaut2017enhanced}, to enhance feature selectivity \cite{poleg2016retinal,sprekeler2017functional}, to modulate neural oscillations \cite{buzsaki2012mechanisms}, and to balance network dynamics \cite{van1996chaos,dehghani2016dynamic}. To understand synaptic mechanisms underlying neuronal computation, it is important to investigate the dynamics of the pure E and I inputs to a neuron via electrophysiological recording techniques. Somatic voltage clamp has become a popular approach to achieve this both \emph{in vitro} and \emph{in vivo} studies over the last thirty years \cite{monier2008vitro}. For instance, voltage clamp has been extensively applied to areas including visual \cite{borg1996voltage,borg1998visual,atallah2012parvalbumin}, auditory \cite{zhang2003topography,wehr2003balanced,wehr2005synaptic,ye2010synaptic}, and prefrontal cortex \cite{shu2003turning,haider2006neocortical}.

To reveal quantitative information of E and I conductances, data collected in voltage clamp mode needs to be further processed to determine the conductance values. In the traditional method, by assuming the neuron as an electrically compact point and the synaptic conductance of this point neuron being \emph{independent} of the injected clamp current, the dynamics of its voltage can be described as \cite{koch2004biophysics}
\begin{equation}
\label{eqn:point}
c\frac{dV}{dt}=-g_{L}(V-\varepsilon_{L})-g_{E}(V-\varepsilon_{E})-g_{I}(V-\varepsilon_{I})+I_{inj},
\end{equation}
where $c$ is the membrane capacitance, $V$ is the membrane potential, $g_{L}$, $g_{E}$ and $g_{I}$ are the leak, E, and I conductances, respectively, $\varepsilon_{L}$, $\varepsilon_{E}$ and $\varepsilon_{I}$ are the corresponding reversal potentials, respectively, and $I_{inj}$ is the externally injected current. Here all potentials are relative to the resting potential. Using the voltage clamp to hold the somatic voltage $V$ at different levels, one can obtain the corresponding synaptic currents $I_{syn}$ and linearly fit an I-V relation at each time point. By casting $g_{E}(\varepsilon_{E}-V)+g_{I}(\varepsilon_{I}-V)$ as $I_{syn}=-kV+b$, the slope
 \begin{equation}
\label{eqn:k1}
k=g_{E}+g_{I}
\end{equation}
is the total conductance (the linear summation of the E and I conductances) and the intercept
 \begin{equation}
\label{eqn:b1}
b=g_{E}\varepsilon_{E}+g_{I}\varepsilon_{I}
\end{equation}
is the reversal current (the weighted summation of the E and I conductances). Therefore, by measuring the slope and the intercept of the I-V relation, one can solve Eqs. (\ref{eqn:k1})-(\ref{eqn:b1}) to obtain the values of $g_{E}$ and $g_{I}$.

Despite the extensive application of voltage clamp to determine E and I conductances, it has yet to address various important issues related to the validity of the above approach. There, an important assumption is that the somatic voltage clamp should uniformly control the membrane potential throughout the entire neuron. However, recent studies \cite{williams2008direct,poleg2011imperfect} have shown there is a space clamp effect \cite{spruston1993voltage} that constrains the voltage clamp to exert only a limited control of the membrane potential across the dendritic arbor. The membrane potential at distal synapses can then deviate greatly from the holding potential. Therefore, the E and I conductances obtained from Eqs. (\ref{eqn:k1}) and (\ref{eqn:b1}) in voltage clamp mode can be distorted significantly from the true local ones.

Despite the space clamp effect, the well-clamped somatic voltage leaves the possibility for the traditional method to measure the effective conductance at the soma, which reflects directly the functional impact of synaptic inputs on action potential initiation and thereby neuronal information processing. However, using theoretical analysis, electrophysiological experiments, and realistic neuron simulations, here we demonstrate that the traditional method conceptually fails to capture the effective conductance because of the neglect of a nonlinear interaction between clamp current at the soma and synaptic currents from dendrites. Consequently, the traditional method incurs substantial error in the measurement of conductance, even giving rise to the unphysical results of negative value as observed in experiments. As a remedy, we devise a method for determining the effective conductance and accordingly verify it in both electrophysiological experiments and realistic neuron simulations, thereby establishing a new framework for elucidating synaptic impact on neuronal computation.

\section*{Effective conductance}
Before the discussion on the inherent problem of the traditional method, we first introduce the concept of effective conductance. Experiments have shown that, although a neuron is not an electrically compact point, the dynamics of its \emph{somatic} membrane potential in response to \emph{somatic} current injection can be well characterized by a point leaky integrator \cite{carandini1996spike,badel2008dynamic}, and the somatic membrane potential can be accurately controlled by the voltage clamp placed at the soma. Based on these facts, it is natural to consider the soma rather than the entire neuron as a point. Accordingly, we introduce the concept of effective conductance at the soma, which is defined, by Ohm's law, as the ratio of the synaptic current $I^{0}_{syn}$ arriving at the soma to the driving force (difference between the reversal potential $\varepsilon$ and the somatic membrane potential $V$) in the presence of either E or I input, i.e.,
\begin{equation}
g^{eff}=\frac{I^{0}_{syn}}{\varepsilon-V}.
\label{eqn:eff_cond}
\end{equation}
It should be stressed that, in order to distinguish from the synaptic current measured using voltage clamp in the traditional method, $I^{0}_{syn}$ with the superscript ``0'' is the synaptic current in the absence of any externally injected current. The effective conductance is an important concept because, as will be demonstrated below, it can be viewed as a proportional indicator of the local postsynaptic conductance on the dendrite. In addition, it reflects directly the functional impact of synaptic inputs on the spike trigger mechanism and thereby neuronal information processing. For example, a strong synaptic input at distal dendrite and a weak synaptic input at proximal dendrite can give rise to a similar magnitude of effective conductance when they arrive at the soma after dendritic filtering and integration, thus inducing a similar somatic response to initiate action potentials and propagate signals. Therefore, from a functional perspective, it could be more meaningful to measure the effective conductance than the local conductance on the dendrite.

On account that the clamp can control the voltage sufficiently well only at the soma and its nearby regions, by applying the somatic voltage clamp, one is purported to measure the effective conductance rather than the local conductance on the dendrite. However, as will be demonstrated below, the conductance determined by the traditional method (using Eqs. \ref{eqn:k1} and \ref{eqn:b1}) has no clear biological interpretation. It is close to neither the local conductance on the dendrite nor the effective conductance at the soma.

\section*{Theoretical analysis}
We first perform a theoretical analysis to illustrate this issue. Assuming a neuron as a linear device, we perform a static transfer resistance analysis \cite{koch2004biophysics,hao2009arithmetic} (see Supplementary Information for details) to obtain a linear relation between the local and effective synaptic conductances
\begin{equation}
\label{eqn:gsei}
g^{eff}_{E}=\frac{K_{ES}}{K_{SS}}g_{E}  \textrm{\ \ \ \ \ \ \ and\ \ \ \ \ \ \ \ } g^{eff}_{I}=\frac{K_{IS}}{K_{SS}}g_{I}
\end{equation}
to the first order accuracy of $g_{E}$ and $g_{I}$, where $g^{eff}_{E}$ and $g^{eff}_{I}$ are the effective E and I conductances respectively, and $g_{E}$ and $g_{I}$ are the corresponding local ones; $K_{ES}$ is the transfer resistance between the location of the E synapse and the soma, $K_{IS}$ is the transfer resistance between the location of the I synapse and the soma, and $K_{SS}$ is the input resistance at the soma. Here the transfer resistance $K_{AB}$ is defined as the ratio of the voltage change in location B to the magnitude of the injected current in location A. Note that transfer resistance is a well-defined property independent of the input when a neuron is within the subthreshold regime, which has been shown in the simulation of our realistic pyramidal neuron model (Fig. 1a). As demonstrated by the linear relation between the local and effective conductances (Eq. \ref{eqn:gsei}), the effective synaptic conductance is indeed a proportional indicator of the local synaptic conductance.

\begin{figure}
\center
\includegraphics[width=80mm]{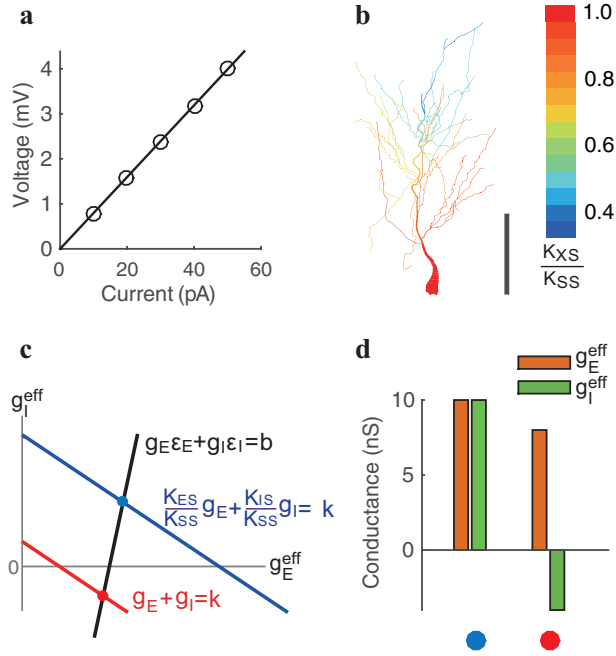}
\caption{Properties of transfer resistance in the determination of effective conductance. \textbf{a}, linear dependence of somatic voltage on the injected current in our realistic neuron model. The slope is the transfer resistance between the input location and the soma. Here the current is injected on the dendrite 50 $\mu$m away from the soma. Model details can be found in Refs. \cite{li2014bilinearity,hao2009arithmetic}. \textbf{b}, spatial profile of the ratio $K_{XS}/K_{SS}$ on the dendritic arbor in stratum radiatum of the realistic neuron. $K_{XS}$ is the transfer resistance between the input location $X$ and the soma $S$, $K_{SS}$ is the input resistance at the soma. Scalar bar indicates 100 $\mu$m. \textbf{c}, an example to illustrate the deficiency of the traditional method in the determination of effective conductance. For a pair of E and I synaptic inputs on the dendrite with $K_{ES}=K_{IS}=0.2K_{SS}$, the corresponding effective conductance values are determined by the intersection between the black and blue lines. Were the traditional approach used to determine the conductance, the intersection between the black and red lines would yield the corresponding conductances. \textbf{d}, the value of the conductances determined in \textbf{c}. Colored circles on the bottom indicate the cases of the intersection points with the corresponding color in \textbf{c}. Note that for the case marked by red, the value of I effective conductance is negative via the traditional method.}
\end{figure}

When a neuron receives both E and I synaptic inputs on the dendrite with its somatic membrane potential clamped at various levels, our analysis further yields a linear relation between synaptic current and voltage in voltage clamp mode (see Supplementary Information for details), i.e., $I_{syn}=-kV+b$,
where
\begin{equation}
\label{eqn:k_local}
k=\Big(\frac{K_{ES}}{K_{SS}}\Big)^2 g_{E}+\Big(\frac{K_{IS}}{K_{SS}}\Big)^2 g_{I}
\end{equation}
and
\begin{equation}
\label{eqn:b_local}
b=\frac{K_{ES}}{K_{SS}} g_{E}\varepsilon_{E}+\frac{K_{IS}}{K_{SS}} g_{I}\varepsilon_{I}
\end{equation}
to the first order accuracy. Alternatively, if we cast Eqs. (\ref{eqn:k_local}-\ref{eqn:b_local}) in terms of effective conductance using Eq. (\ref{eqn:gsei}), we can obtain
\begin{equation}
\label{eqn:k2}
k=\frac{K_{ES}}{K_{SS}}g^{eff}_{E}+\frac{K_{IS}}{K_{SS}}g^{eff}_{I}
\end{equation}
and
\begin{equation}
\label{eqn:b2}
b=g^{eff}_{E}\varepsilon_{E}+g^{eff}_{I}\varepsilon_{I}
\end{equation}
to the first order accuracy. Therefore, to determine the local conductance, one needs to solve $g_E$ and $g_I$ from Eqs. (\ref{eqn:k_local})-(\ref{eqn:b_local}); while, to determine the effective conductance, one needs to solve $g_E^{eff}$ and $g_I^{eff}$ from Eqs. (\ref{eqn:k2})-(\ref{eqn:b2}). Clearly, in contrast to Eq. (\ref{eqn:k1}) in the traditional method, the slope $k$ of the I-V relation in Eq. (\ref{eqn:k_local}) or Eq. (\ref{eqn:k2}) is neither the total effective conductance nor the total local conductance. In this sense, the conductance determined by the traditional method using Eqs. (\ref{eqn:k1})-(\ref{eqn:b1}) is neither the local conductance nor the effective conductance. We can further show that there does not exist a transform between the local conductance and the conductance determined by the traditional method (see Supplementary Information), thus the conductance determined by the traditional method has no clear biological interpretation. The prefactors $\frac{K_{ES}}{K_{SS}}$ and $\frac{K_{IS}}{K_{SS}}$ in Eqs. (\ref{eqn:k_local})-(\ref{eqn:k2}) arise from the nonlinear interaction between the injected current at the soma and the synaptic current from the dendrite (see Supplementary Information for details). Only when the E and I inputs are given at the soma will the prefactors vanish. In this particular limit, the local and effective conductances become identical (Eq. \ref{eqn:gsei}). In addition, Eqs. (\ref{eqn:k_local})-(\ref{eqn:b_local}) and (\ref{eqn:k2})-(\ref{eqn:b2}) further reduce to Eqs. (\ref{eqn:k1})-(\ref{eqn:b1}), which enables one to use the traditional method to determine the local or effective conductance. However, in general, these prefactors cannot be naively assumed to be unity (i.e., no nonlinear interaction) since they can distort significantly the determination of conductance, as will be demonstrated below.

As it is challenging in experiment to elicit inputs which are spatially broadly distributed on the distal dendrite, we resort to numerical simulations using the realistic pyramidal neuron model to investigate the spatial dependence of the prefactors $\frac{K_{ES}}{K_{SS}}$ and $\frac{K_{IS}}{K_{SS}}$. Our numerical result shows that the prefactors across the entire dendritic tree decay from unity to zero rapidly with the increase of the distance between the synaptic input sites and the soma (Fig. 1b). Therefore, if one attempts to determine the effective conductance using the traditional method based on Eqs. (\ref{eqn:k1})-(\ref{eqn:b1}) rather than Eqs. (\ref{eqn:k2})-(\ref{eqn:b2}), the errors can become prominent. For instance, when the E and I inputs are given at the distal dendrite where the prefactors are small, the value of conductance could vanish when determined by the traditional method. In addition, in our analysis, when the prefactors become sufficiently small, a negative conductance can arise via the traditional method (Fig. 1c-1d). This possibly explains why a negative conductance was observed in early experiments \cite{williams2008direct}. Our theoretical prediction is particularly of note that the measurement of I conductance is distorted more significantly than E conductance by the traditional method --- as a consequence of the ratio of measurement error between the E and I conductances being proportional to the ratio between the I reversal potential $\varepsilon_I$ (e.g., $-10$ mV relative to the resting potential) and the E reversal potential $\varepsilon_E$ (e.g., 70 mV relative to the resting potential), i.e., $\Delta g_{E}:\Delta g_{I}=-\varepsilon_I:\varepsilon_E$ (see Fig. 1c and Supplementary Information).

In principle, the error of the traditional method could be eliminated by measuring the value of the prefactors associated with input locations. However, for a neuron receiving a large number of spatially broadly distributed synaptic inputs, it remains difficult to have all \emph{a priori} information about the transfer resistances between the synaptic input sites and the soma, thus hampering the recovery of the E and I conductances from Eqs. (\ref{eqn:k2})-(\ref{eqn:b2}) directly. From our theoretical analysis, we note that the intercept $b$ in Eq. (\ref{eqn:b2}) possesses the form of effective synaptic reversal current without the information of transfer resistances. Therefore, we propose to recover the effective E and I conductances only from the intercept information by varying either the E or I reversal potential at various levels. For example, we can vary the I reversal potential from $\varepsilon_{I}$ to $\varepsilon^{'}_{I}$ to obtain a second intercept equation
\begin{equation}
\label{eqn:b3}
b^{'}=g^{eff}_{E}\varepsilon_{E}+g^{eff}_{I}\varepsilon^{'}_{I},
\end{equation}
and then the effective E and I conductances can be obtained from Eqs. (\ref{eqn:b2}) and (\ref{eqn:b3}). In physiological experiment, to change reversal potential, one needs to effect a change of the extracellular or intracellular fluid environment. From now on, we refer to this method based on Eqs. (\ref{eqn:b2})-(\ref{eqn:b3}) as the intercept method (IM), and the traditional method based on Eqs. (\ref{eqn:k1})-(\ref{eqn:b1}) as the slope-and-intercept method (SIM).
\section*{Electrophysiological experiments}
To confirm our theoretical results above, we perform electrophysiological experiment to demonstrate the interaction between synaptic current and injected clamp current. In the experiment (see Methods for details), we record 7 rat CA1 pyramidal neurons using somatic voltage clamp. The resting potential of each neuron ranges from $-57$ mV to $-68$ mV. E and I synaptic inputs are given via a dynamic clamp at the location on the dendrite about 100 $\mu$m away from the soma. The absolute E and I reversal potentials are set as $E_{AMPA}=0$ mV and $E_{GABA_A}=-70$ mV, and the relative reversal potentials $\varepsilon_E$ and $\varepsilon_I$ can be determined by subtracting the resting potential from $E_{AMPA}$ and $E_{GABA_A}$ respectively. The local input synaptic conductances through the dynamic clamp take the form of a difference of two exponential functions whose time constants were derived from voltage traces in experiment \cite{hao2009arithmetic}, with rise time constant 5 ms (6 ms) and decay time constant 7.8 ms (18ms) for E (I) conductance. The peak amplitude of E and I synaptic conductances ranges from 2 nS to 5 nS and 3 nS to 6 nS, respectively. For each pyramidal neuron, we clamp the voltage at the soma with five levels from $-50$ mV to $-90$ mV with an increment of 10 mV (Fig. 2a, 2c). For an individual E input given on the dendrite via dynamic clamp (Fig. 2a), we can then record five excitatory postsynaptic current (EPSC) traces $I^{inj}_{syn}$ at the soma corresponding to the five holding voltage levels, and determine the corresponding E conductance traces using $g^{inj}_{E}=I^{inj}_{syn}/(\varepsilon_{E}-V)$. Here $\varepsilon_E$ and $V$ are reversal and holding potentials relative to the resting potential, and $I^{inj}_{syn}$ is the synaptic current which is the current increment from the baseline injected current that is used to clamp the neuron to a steady state of voltage (see Supplementary Information). The superscript ``$inj$'' in the notations $g^{inj}_{E}$ and $I^{inj}_{syn}$ emphasizes the fact that they are determined \emph{in the presence of} the injected clamp current. We obtain the final profile of the resulting conductance as a function of a difference of two exponentials using least square fitting. Fig. 2b shows that five E conductances $g^{inj}_{E}$ thus obtained are not identical with disparity between these conductances well beyond recording statistical fluctuations. The dependence of the conductance value of $g^{inj}_{E}$ on the clamp voltage as shown in Fig. 2b contradicts the assumption of the traditional SIM that the synaptic conductance is independent of injected current as described in Eq. (\ref{eqn:point}). It confirms our result that the synaptic current from the dendrite and the injected current on the soma cannot be linearly summed as a consequence of their interaction with each other. The difference between the I conductances $g^{inj}_{I}$ estimated from different voltage clamp levels is more prominent than the E case (Fig. 2c, 2d). The voltage dependence of the I conductance $g^{inj}_{I}$ can be highlighted by the following limiting case. When the soma is clamped at the I reversal potential $E_{GABA_A}=-70$ mV, the value of the I conductance $g^{inj}_{I}$ would become unphysically infinity because the denominator in the expression $g^{inj}_{I}=I^{inj}_{syn}/(\varepsilon_{I}-V)$ vanishes (such a case is not displayed in Fig. 2d).

There is further evidence demonstrating the interaction between synaptic current and injected clamp current as observed in the experiment. Were the synaptic and injected currents linearly summable (Eq. \ref{eqn:point}), then the conductances obtained from the slope and those from the intercept of the I-V relation would be identical through SIM. However, this turns out not to be the case in our experimental observation. To be specific, given an individual E or I input on the dendrite, we can obtain a linear I-V relationship between the voltage and the synaptic current $I^{inj}_{syn}$ at each time point after the onset of the stimulus. An example of the I-V relationship at the time 10 ms after the stimulus onset is shown in Fig. 2e. Upon casting $g(\varepsilon-V)$ as $I^{inj}_{syn}$ and following Eqs. (\ref{eqn:k1}) and (\ref{eqn:b1}) for the case of only a purely E or I input, we obtain the ratio of the conductance value estimated from the slope to that from the intercept. This ratio deviates greatly from unity --- the expected result obtained by SIM. It is nearly a constant and is independent of conductance amplitude (Fig. 2f). The value of ratio for E inputs is nearly identical to that for I inputs when the E and I inputs are given at the same dendritic location (Fig. 2f). This observation is consistent with our theoretical prediction from Eqs. (\ref{eqn:k2}) and (\ref{eqn:b2}) for the case of a purely E or I input, for which the above ratio is the same as the ratio of the effective conductance $g^{eff}_{E}$ ($g^{eff}_{I}$) at the soma to the local conductance $g_{E}$ ($g_{I}$) on the dendrite (Eq. \ref{eqn:gsei}). Based on our analysis, the conductance obtained from the intercept of the I-V relation is the true effective conductance, i.e., the conductance at the soma induced by a synaptic input on the dendrite \emph{in the absence of} the injected current. Therefore, we choose the values of the E and I conductances estimated from the intercepts for the case of only pure E or I inputs as the \emph{reference conductances} to evaluate IM below. We note in passing that the effective reference conductances determined in this way are more accurate than those determined directly from Eq. (\ref{eqn:point}) in the absence of voltage clamp for we can avoid taking time derivative of noisy experimental voltage data --- a procedure that would introduce large numerical errors. In our realistic neuron simulation to corroborate our experimental results below, however, we can use Eq. (\ref{eqn:point}) in the absence of voltage clamp to determine the effective conductance since the numerical simulation is sufficiently accurate for obtaining the time derivative of voltage.

\begin{figure}
\center
\includegraphics[width=120mm]{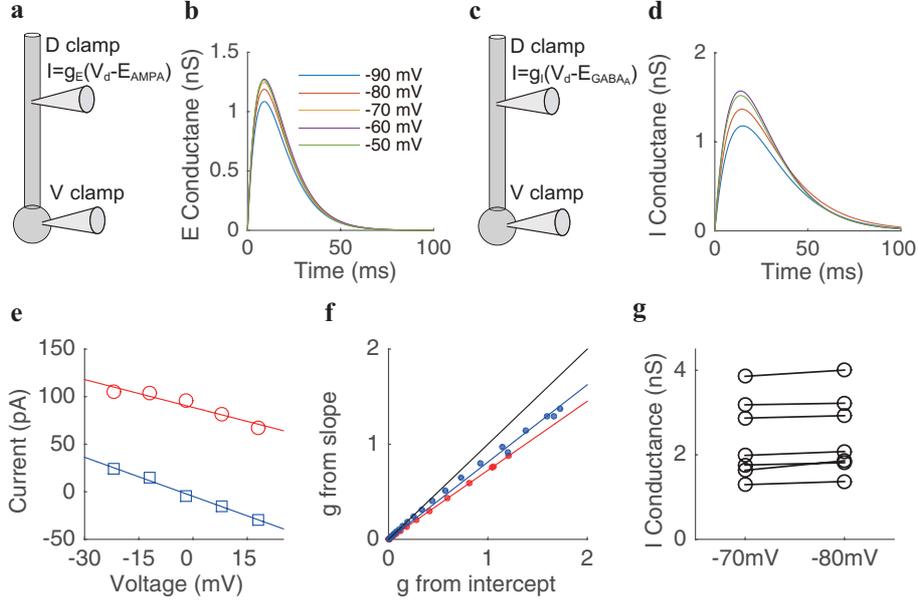}
\caption{Determination of individual effective conductance in experiments. \textbf{a}, schematic diagram of the recording configuration when a pyramidal neuron receives an individual E input. A voltage clamp is made at the soma and a dynamic clamp is made on the dendritic trunk about 100 $\mu$m away from the soma. Somatic voltage is clamped from $-90$ mV to $-50$ mV when an individual E input is given by dynamic clamp on the dendrite. The local E conductance takes the form proportional to $e^{-t/5}-e^{-t/7.8}$ (time unit: ms) with a peak amplitude of 2.5 nS. \textbf{b}, $g^{inj}_{E}$ obtained as $I^{inj}_{syn}/(\varepsilon_{E}-V)$ by least square fitting of a double exponential function.  \textbf{c}, schematic diagram of the recording configuration when a pyramidal neuron receives an individual I input --- same as in \textbf{a} except that the local I conductance takes the form proportional to $e^{-t/6}-e^{-t/18}$ (time unit: ms) with a peak amplitude of 4 nS. \textbf{d}, $g^{inj}_{I}$ obtained as $I^{inj}_{syn}/(\varepsilon_{I}-V)$ by least square fitting of a double exponential function (cf. \textbf{b} for color coding). Not shown is the case for the clamp voltage at the absolute reversal potential of $-70$ mV (see text). \textbf{e}, I-V relations corresponding to E (red) and I (blue) inputs under the five holding voltage levels, respectively. The data are collected at the time 10 ms after the onset of the stimulus. The value of the voltage shown in the abscissa is relative to the resting potential. \textbf{f}, the relation between the conductance determined from the slope (Eq. \ref{eqn:k1}) and the intercept (Eq. \ref{eqn:b1}). The E and I conductances are marked by red and blue color, respectively. Each dot corresponds to the conductance value at one time point. The red and blue lines are corresponding linear fitting. For reference, the black reference line has a slope of unity.  \textbf{g}, the independence of the I conductance value on the change of the I reversal potential. The circles on the left and right columns are the peak I conductances determined with the I reversal potential $E_{GABA_A}=-70$ mV and $E_{GABA_A}=-80$ mV, respectively. The solid lines link the conductances determined at the same neuron. The mean relative change of the conductance under different reversal potentials is within $5\%$.}
\end{figure}

We next perform experiment to demonstrate the validity of IM by contrasting its error with that of SIM. Because we need to vary the reversal potential to a different value in IM, we have first verified that the value of the effective conductance is nearly independent of the reversal potential value (Fig. 2g). Next, we proceed to determine the effective conductance. Given an individual E pulse input at a dendritic location about 100 $\mu$m away from the soma and placing the voltage clamp at the soma, we can record a set of synaptic current $I^{inj}_{syn}$ under five holding voltages from $-50$ mV to $-90$ mV. We then determine the effective E conductance from the intercept of the I-V relation at each time point. A similar procedure is carried out separately for the effective I conductance. A pair of measured effective E and I conductances determined in this way is displayed in Fig. 3c as the reference values (solid curves), against which we evaluate the performance of IM and SIM. Next, with a voltage clamp placed at the soma, we simultaneously elicit the E and I pulse inputs same as for the reference ones, i.e., input at the same dendritic location with the same strength (Fig. 3a). Five total synaptic currents $I^{inj}_{syn}$ at the soma are obtained under five holding voltages from $-50$ mV to $-90$ mV. We then also observe a linear relation between the synaptic current and the membrane potential at each time point. An example of the I-V relation at the time 10 ms after the onset of the stimulus is shown in Fig. 3b. Finally, we determine a pair of values of the E and I conductance pulses from the linear I-V relation by using SIM. Meanwhile, by changing the I reversal potential $E_{GABA_A}$ from $-70$ mV to $-80$ mV and repeating the above procedure (Fig. 3b), a pair of alternative values of E and I conductances can be obtained using IM. By comparing the values of the conductance pulses measured by the two methods with those of the reference conductance pulses in Fig. 3c, we observe that the conductance estimated by IM nearly overlaps with the true conductance, whereas the conductance estimated by the traditional SIM deviates greatly from the true conductance. As shown in Fig. 3d, across 7 pyramidal neurons, the effective conductance measured by IM has a relatively small error on average, with a relative error of peak amplitude ranging from $0.6\%$ to $15.6\%$ for E conductance and from $3.0\%$ to $14.3\%$ for I conductance. In contrast, the conductance measured by SIM yields a large relative error of peak amplitude as great as from $13.1\%$ to $40.0\%$ for E conductance and from $10.1\%$ to $44.9\%$ for I conductance. According to our theoretical analysis, the error in SIM is caused by failing in taking into account the nonlinear interaction between the synaptic current from the dendrite and the injected current at the soma --- thus missing the prefactors $K_{ES}/K_{SS}$ and $K_{IS}/K_{SS}$ in Eq. (\ref{eqn:k2}), whose strength has a sensitive dependence on the dendritic location (Fig. 1b). In our experiments, we have given inputs on the proximal dendrite only about 100 $\mu$m away from soma, nevertheless, the error has already reached $\sim40\%$ (Fig. 3d). For a synaptic input location further towards the distal dendrite, the error is expected to become substantially larger.

\begin{figure}
\center
\includegraphics[width=80mm]{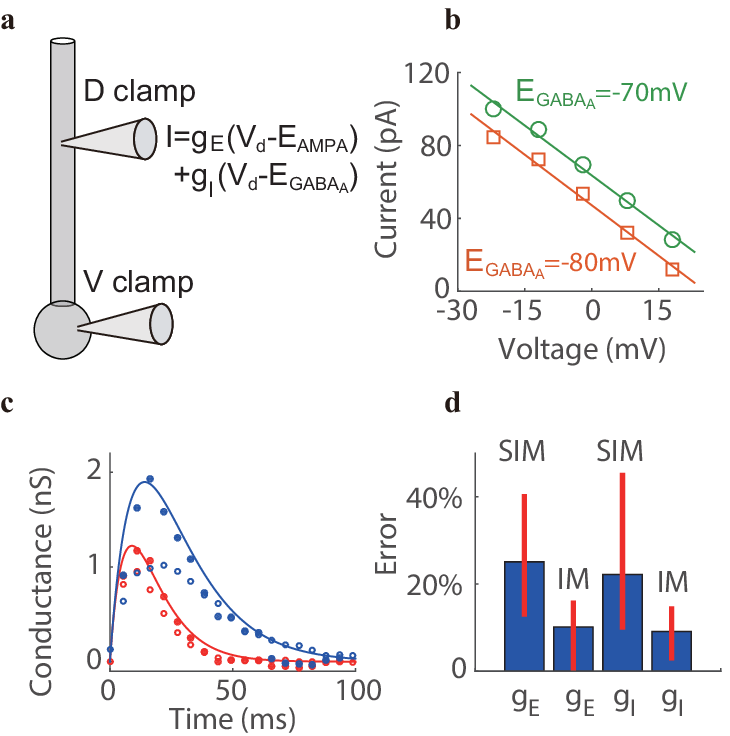}
\caption{Determination of a pair of effective conductances in experiment. \textbf{a}, schematic diagram of the recording configuration when a pyramidal neuron receives a pair of E and I inputs. A voltage clamp is made at the soma and a dynamic clamp is made on the dendritic trunk about 100 $\mu$m away from the soma. Somatic voltage is clamped from $-90$ mV to $-50$ mV when both E and I inputs are given simultaneously by the dynamic clamp on the dendrite. The input parameters are the same as in Fig. 2. \textbf{b}, I-V relations obtained after applying the voltage clamp with the I reversal potentials $E_{GABA_A}=-70$ mV (green) and $E_{GABA_A}=-80$ mV (orange), respectively. The data are collected at the time 10 ms after the onset of the stimulus. The voltage value shown in the abscissa is relative to the resting potential. \textbf{c}, a pair of E (red) and I (blue) conductances determined by SIM (open circle) and IM (solid circle). Solid curves are the reference conductances measured by given an individual E or I input separately. \textbf{d}, the peak-amplitude relative error of the E and I conductances determined by IM and SIM across 7 neurons. The thick blue bar indicates the mean value of the peak amplitude error and the thin red bar indicates the range of the error.}
\end{figure}
\section*{Realistic neuron simulations}
As it is a rather challenging experimental task to elicit inputs on multiple dendritic locations in general, and on the distal dendrite in particular, we turn to realistic neuron simulations to assess the performance of IM and SIM for distal inputs or spatiotemporally broadly distributed multiple inputs on the dendrite. In our realistic pyramidal neuron model (adapted from that in Refs. \cite{li2014bilinearity,hao2009arithmetic}), the resting potential is set to $V_{r}=-70$ mV, and the absolute E and I reversal potentials are initially set to $E_{AMPA}=0$ mV, $E_{GABA_A}=-80$ mV. The relative reversal potentials are then determined as $\varepsilon_{E}=70$ mV and $\varepsilon_{I}=-10$ mV. As above, we first determine the reference E and I conductances by giving the neuron an individual E and I input separately. For an individual E pulse input at a dendritic location about 350 $\mu$m away from the soma but without the injected clamp current at the soma, we can numerically record the corresponding EPSP at the soma and invoke Eq. (\ref{eqn:eff_cond}) to determine the value of the effective E conductance pulse from the point-neuron model (for which we set $I_{inj}=0$, $g_{I}=0$ in Eq. \ref{eqn:point}). A similar procedure can be carried out for the effective I conductance pulse in response to an individual I pulse input at a dendritic location about 300 $\mu$m away from the soma (again, in the absence of injected current at the soma). In the simulation, the experimental result is also confirmed that the value of the effective E or I conductance is nearly identical under different synaptic reversal potentials (Fig. 4a). Under simultaneous E and I inputs, the application of voltage clamp gives rise to an I-V relation at each time point as in experiment. An additional I-V relation at the same time point after the onset of the stimulus results from altering the I reversal potential $E_{GABA_A}$ from $-80$ mV to $-90$ mV (Fig. 4b). As shown in Fig. 4c, the conductances measured using IM have a small relative error compared with the corresponding reference values, with a maximum error of $6\%$ for both E and I conductances in the peak amplitude. In contrast, those determined using SIM yield an error as large as $31\%$ for E conductance and $75\%$ for I conductance.

To model the situation \emph{in vivo}, we distribute 15 E inputs and 5 I inputs across the entire dendritic tree of the pyramidal neuron (Fig. 4d). At each synaptic location, the arrival time of each input is randomly selected between 0 ms and 1000 ms with input rate of 100 Hz. We use the identical input for both the measurement of time evolution of the effective conductance as reference without the voltage clamp (using Eq. \ref{eqn:point}) as that of the conductances with the voltage clamp. Comparison of the values of conductance measured by the IM and SIM methods with the reference conductance in Fig. 4e demonstrates that the effective E or I conductance estimated by our IM is in good agreement with the true effective conductance, except for the moments when the neuron generates action potentials \cite{guillamon2006estimation} in the neuronal dynamics without voltage clamp. Meanwhile, the conductance estimated by SIM deviates greatly from the true one in general, and particularly substantial for the inhibitory case. In the subthreshold regime, the conductances measured by IM incur a relatively small error with time averaged relative error of $26\%$ for E conductance and $8\%$ for I conductance, whereas those determined using SIM yield a time averaged relative error as large as $47\%$ for E conductance and $89\%$ for I conductance. In this simulation, while the true I conductance is substantially larger than the true E conductance, the I conductance estimated by SIM turns out to be substantially smaller than the E conductance. It is important to stress that the value of the I conductance could even become negative, thus demonstrating the severe deficiency of SIM. We note that in Fig. 4c the tail of the reference E conductance also becomes slightly negative, which arises from the repolarization of the membrane potential before relaxing to its resting state. This phenomenon has also been observed in experiments \cite{hao2009arithmetic} and is attributed to the activity of voltage-gated ion channels, which have not been taken into account in the simple point neuron model (Eq. \ref{eqn:point}). Unlike the case of the E conductance, the negative I conductance determined by SIM in Fig. 4e originates from the deficiency of the method itself instead of active channels. Contrary to the case in Fig. 4c, the negative value of conductance in Fig. 4e (away from the moments of the action potentials) is only observed for the inhibitory one, for which its reference conductance always stays at a positive level.

We now address the question of how the error of the two methods depends on the input locations on the dendrite. For a pair of synaptic inputs at various locations on the dendrite, our simulation shows that IM can control the error within $10\%$ even for the distal inputs (Figs. 4f and 4h), whereas the error of SIM increases rapidly to $100\%$ as the input location moves away from the soma to the distal dendrite (Figs. 4g and 4i). In some remote distal dendritic sites --- greater than $400$ $\mu$m away from the soma, the estimated I conductance can also become negative (Fig. 4i), accentuating the SIM deficiency.

A further validation of IM is shown in Fig. 5. For the same pair of transient E and I inputs in Fig. 4c, on the one hand, we can measure the slope and the intercept of the I-V relation using voltage clamp when the E and I inputs are given simultaneously; on the other hand, we can determine the effective E and I reference conductances $g^{eff}_{E}$ and $g^{eff}_{I}$ when the E and I inputs are given separately and then reconstruct the total conductance as $g^{eff}_{E}+g^{eff}_{I}$ and the effective reversal current as $g^{eff}_{E}\varepsilon_{E}+g^{eff}_{I}\varepsilon_{I}$. Fig. 5 shows that the effective reversal current overlaps well with the intercept of the I-V relation, while the violation of SIM is instantiated by a rather substantial difference between the total effective conductance and the slope of the I-V relation.

\begin{figure}
\center
\includegraphics[width=160mm]{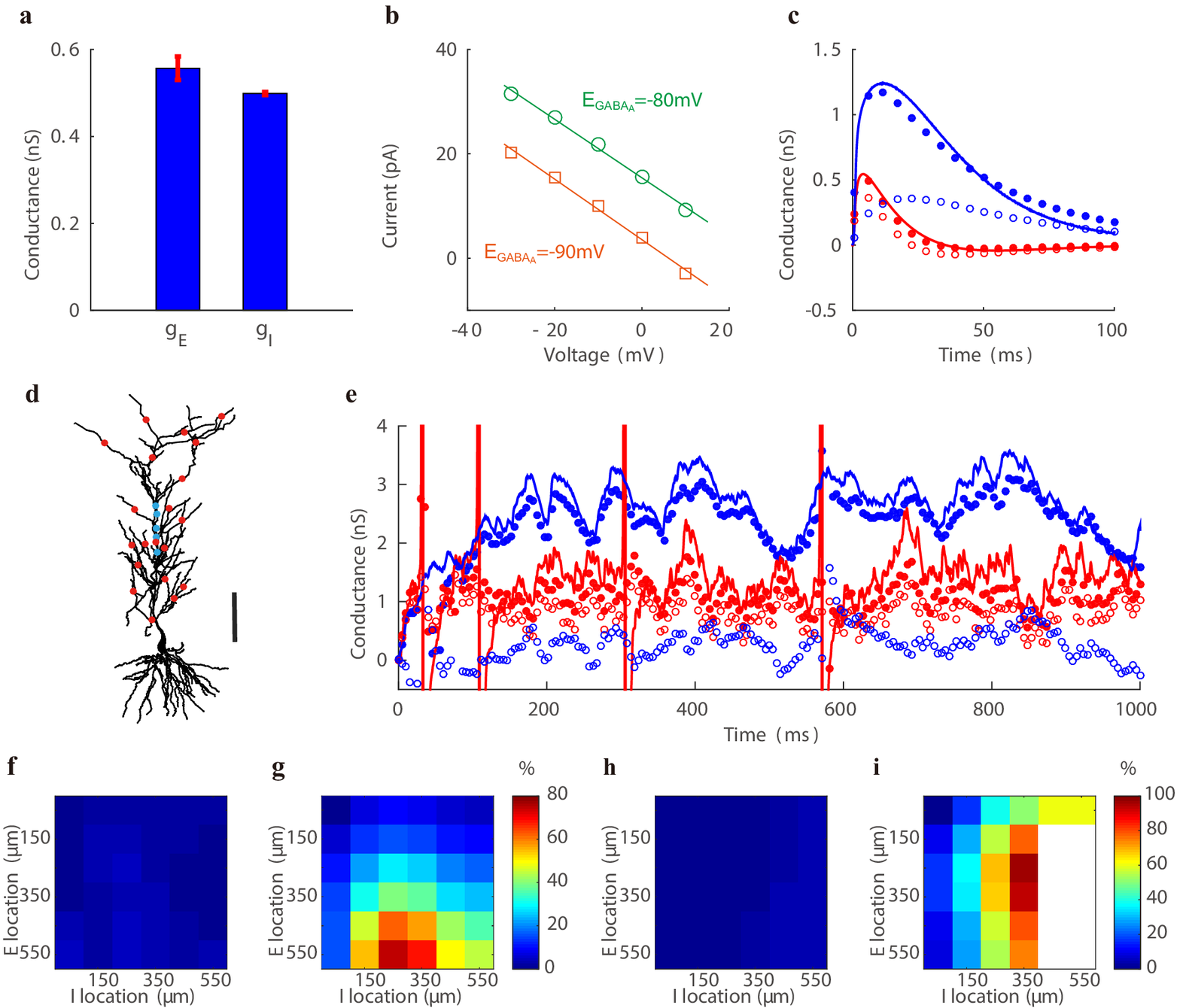}
\linespread{1}
\caption{Determination of effective conductances in realistic neuron simulations. \textbf{a}, independence of the E and I conductances on the change of the reversal potentials. The E reversal potential $E_{AMPA}$ varies from $-50$ mV to 50 mV with ten even increments, and the I reversal potential $E_{GABA_A}$ varies from $-70$ mV to $-90$ mV with five even increments. The thick blue bar indicates the mean and the thin red bar indicates one standard deviation. The ratio of standard deviation to mean for both conductance values is within $5\%$. \textbf{b}, Two I-V relations obtained under somatic voltage clamp mode with the I reversal potential $E_{GABA_A}$ changing from $-80$ mV (green) to $-90$ mV (orange). The data are collected at the time 15 ms after the onset of the stimulus. The voltage value shown in the abscissa is relative to the resting potential. \textbf{c}, a pair of E (red) and I (blue) effective conductances determined by SIM (open circle) and IM (solid circle). The E and I inputs are given simultaneously on the dendritic trunk about 350 $\mu$m and 300 $\mu$m away from the soma, respectively. Solid curves are the reference conductances obtained under the same individual E or I input given separately. \textbf{d}, the spatial distribution of multiple E (red) and I (blue) inputs on the dendrite. Bar indicates 100 $\mu$m. \textbf{e}, E (red) and I (blue) effective conductances determined by SIM (open circle) and IM (solid circle) when the neuron receives multiple inputs. Solid curves are the reference conductances obtained by the same  E or I inputs given separately. The input locations are shown in \textbf{d} and the input times at each location are uniformly distributed from 0 ms to 1000 ms with a rate of 100 Hz. Red vertical bars indicate spike events in the absence of voltage clamp. \textbf{f-i}, spatial dependence of the relative error for the E and I conductance measurement. Here, the locations for a pair of E and I input of constant conductances are scanned across the dendrite. The location distance is measured from the soma. \textbf{f-g} are the error of E conductance measured by IM and SIM, respectively, with a common color bar. \textbf{h-i} are the error of I conductance measured by IM and SIM, respectively, with a shared color bar to indicate the percentage of error. In \textbf{i}, the large white area, which corresponds to negative conductance values, again illustrates the failure of the traditional method.}
\end{figure}

\begin{figure}
\center
\includegraphics[width=80mm]{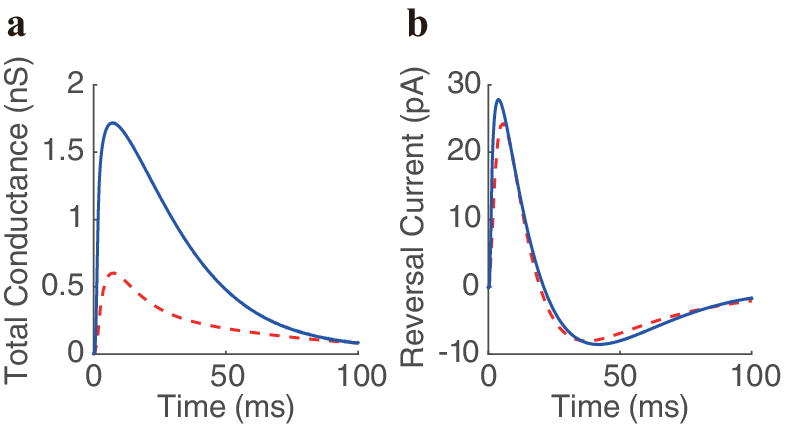}
\caption{Direct validation of IM from realistic neuron simulations. Given the same pair of the synaptic inputs simultaneously as shown in Fig. 4c to obtain reference effective E and I conductances $g^{eff}_{E}$ and $g^{eff}_{I}$, an I-V relation at a particular moment can be obtained by holding the somatic voltage at different levels. \textbf{a}, the deviation of the slope of the I-V relation (red dash curve) from the total conductance (blue curve) calculated as $g^{eff}_{E}+g^{eff}_{I}$. \textbf{b}, good agreement between the intercept of the I-V relation (red dash curve) and the reversal current (blue curve) calculated as $g^{eff}_{E}\varepsilon_{E}+g^{eff}_{I}\varepsilon_{I}$.}
\end{figure}

\section*{Discussion}
To extract E and I conductances, many previous works assumed that a neuron with its complex dendritic arbor is of a single electrically compact compartment. Hence, the somatic voltage clamp is deemed to uniformly control the membrane potential throughout the entire neuron. In the present work, we have explored a better approximation by viewing the soma rather than the entire neuron as an electrically compact point (Eq. \ref{eqn:point}) so as to deploy a nearly perfectly clamped voltage at the soma to measure effective E and I conductances at the soma. Importantly, by taking into account the effect of the nonlinear interaction between synaptic currents from the dendrite and the injected current at the soma, we reveal the major deficiency of SIM, that is lack of clear biological meaning of its measured conductance value and its related substantial error with respect to local or effective conductance.

Under the new framework, we have proposed the concept of the effective conductance. It reflects intrinsically the effect of active ion channels along the dendrite and the filtering property of the dendrite. Furthermore, the effective conductance is a functionally important quantity because it is strongly correlated to the local postsynaptic conductance at the dendrite, and it can gauge more directly than the local conductance the functional impact of synaptic inputs on the subthreshold dynamics and the spike trigger mechanism. We then have proposed a new method --- the intercept method --- to measure the effective conductance accurately and demonstrated that IM produces rather good measurement of conductance values using rat hippocampal CA1 pyramidal neurons and a biologically realistic model neuron. As has been noted above, there is the space clamp effect \cite{williams2008direct,poleg2011imperfect} which limits the control of the voltage clamp on the membrane potential across the entire dendritic arbor, thus potentially impeding the quantitative understanding of synaptic physiology. The notion of the effective conductance, however, obviates the issue of how well one can control the voltage clamp on the dendrite since we only need to control the clamped voltage well at the soma in our method. IM can be applied in many settings in neurophysiological studies involving the E-I interaction, for instance, to understand synaptic mechanisms of sensory processing, the origin of neuronal oscillations, and the balanced nature of excitation and inhibition. From our theoretical analysis, IM is applicable for a large class of neurons in a wide physiological regime for which an approximate linear input-output relation holds \cite{cash1998input,cash1999linear}. Incidentally, we note that IM can also be applied under a current clamp \cite{priebe2005direction,wilent2004synaptic}.

Finally, it is worthwhile to comment that our analysis is accurate only to the first order approximation of the effective conductance, and higher order corrections may also contribute to the conductance value; In addition, our analysis is based on a simple point model of the soma. The dendritic integration of synaptic inputs can potentially lead to a more complicated form of a point-neuron model of the soma \cite{zhou2013phenomenological}. It is important to address these issues in the future study for a quantitative understanding of the synaptic dynamics of neurons with high precision.

\section*{Methods}
\subsection*{Slice electrophysiology}

The preparation of acute hippocampal slices (350 $\mu$m thick) from Sprague Dawley rats of postnatal days 15-20 followed a method described in our previous study \cite{hao2009arithmetic}. The animal experimental protocol was approved by the Animal Use and Care Committee of State Key Laboratory of Cognitive Neuroscience \& Learning at Beijing Normal University. In brief, rats were deeply anesthetized by i.p. injection of pentobarbital (30 mg/kg), and the brain was quickly dissected and then incubated in the ice-cold artificial cerebrospinal fluid (aCSF), which was oxygenated with $95\%$ O$_{2}$ / $5\%$ CO$_{2}$. Coronal hippocampal slices were sectioned with vibratome (VT1200, Leica) and incubated in oxygenated aCSF at 34 $^\circ$C for 30 min, followed by an incubation at 20-22 $^\circ$C till the use for the electrophysiological recording. The aCSF contained (in mM) 125 NaCl, 3 KCl, 2 CaCl$_2$, 2 MgSO$_4$, 1.25 NaH$_2$PO$_4$, 1.3 sodium ascorbate, 0.6 sodium pyruvate, 26 NaHCO$_3$, and 11 D-glucose (pH 7.4 bubbled with $95\%$ O$_2$ / $5\%$ CO$_2$).

Whole-cell recording was made on the hippocampal CA1 pyramidal cell (PC) in slices in a chamber perfused with the aCSF solution (2 ml/min; 30-32$^\circ$C), under an Olympus upright microscope (BX51WI) that was equipped with the differential interference contrast (DIC) and fluorescence optics as well as an infrared camera (IR-1000E, DAGE-MTI). The borosilicate-glass micropipettes were pulled by a Sutter puller (P-1000) and filled by an internal solution containing (in mM) 145 K-gluconate, 5 KCl, 10 HEPES, 10 disodium phosphocreatine, 4 Mg-ATP, 0.3 Na-GTP and 0.2 EGTA (pH 7.3, 295 mOsm). Simultaneous recordings from the cell body and dendrite of a PC followed a procedure reported previously \cite{davie2006dendritic}, in which whole cell recording on the soma was first made using a micropipette (3-5 M$\Omega$; with 20 $\mu$M Alexa Fluor 488, InvitroGene), followed by another recording on Alexa Flour 488 (green)-labeled apical dendritic arbor at position $\sim$100 $\mu$m away from the soma with a micropipette (10-15 M$\Omega$, filled with the internal solution without Alexa Fluor 488). The serial resistance was compensated by $>90\%$ using the built-in function of the amplifier MultiClamp 700B (Molecular Devices). Holding potentials of recorded cells were corrected for a calculated liquid junction potential \cite{barry1994jpcalc} of $\sim$15 mV. In the dynamic clamp recording experiments, either AMPA type glutamate receptor-mediated excitatory conductance or GABA$_\textrm{A}$ receptor-mediated inhibitory conductance was intracellularly injected to the recorded PCs through the whole-cell recording pipette, using the built-in dynamic-clamp function of a 1401 Power3 digitizer (CED) and the Spike2 software (v5.08; CED). Kinetics of AMPA or GABA$_\textrm{A}$ receptor conductance were in the form of two exponential functions with different rise/decay time constants: 5/7.8 ms for AMPA conductance; 6/18 ms for GABA$_\textrm{A}$ conductance. Their respective reversal potentials, $E_{AMPA}$ and $E_{GABA_A}$, were set as 0 mV and $-70$ mV. Membrane voltage or current signals were amplified with a MultiClamp 700B amplifier (Molecular Devices), filtered at10 KHz (low-pass), digitalized by an analog-digital converter (1401 Power3, CED) at 50 KHz, and then acquired by the Spike2 software into a computer for further analysis.

\section*{Acknowledgments}
This work was supported by NYU Abu Dhabi Institute G1301 (S.L., D.Z., and D.C.), NSFC-11671259, NSFC-11722107, NSFC-91630208, Shanghai Rising-Star Program-15QA1402600 (D.Z.), NSFC 31571071, NSF DMS-1009575 (D.C.), Shanghai 14JC1403800, Shanghai 15JC1400104, SJTU-UM Collaborative Research Program (D.Z. and D.C.), the State Key Research Program of China 2011CBA00404 (X-h. Z.)\\

\noindent Correspondence and requests for materials should be addressed to X-h.Z. (xhzhang@bnu.edu.cn), or D. Z. (zdz@sjtu.edu.cn), or D.C. (cai@cims.nyu.edu).


\begin{small}
\bibliographystyle{plain}
\bibliography{arXiv_template}
\end{small}

\section*{Supplementary Information}
\noindent Here we generalize the static two-port analysis \cite{Koch2004} to determine theoretically the E and I conductances received simultaneously by a target neuron with spatial structure. As a static analysis, the synaptic inputs are independent of time. The case of time-dependent synaptic input is illustrated in both simulations of a biologically realistic pyramidal neuron and electrophysiological experiments of rat CA1 pyramidal neurons in the main text.\\

\subsection*{Derivation of Effective Conductance}
\noindent As proposed in the main text, the effective conductance plays an important role in understanding of synaptic inputs and their effects on the soma. Here, we derive the relationship between local synaptic conductance induced at synapses on the dendrite and effective conductance measured at the soma. If a neuron receives an E input on the dendrite, the local synaptic current on the dendrite can be characterized by Ohm's law,

\begin{equation}
\label{eqn:1}
I_{E}=g_{E}(\varepsilon_{E}-V_{E}),
\end{equation}

\noindent where $g_{E}$ is the local E conductance at the synapse, $\varepsilon_{E}$ is the E reversal potential, and $V_{E}$ is the local membrane potential at the synapse. Unless otherwise specified, all potentials are relative to the resting potential. \\

\noindent Based on Ohm's law, the local membrane potential $V_{E}$ can be computed by
\begin{equation}
\label{eqn:2}
V_{E}=K_{EE}I_{E},
\end{equation}

\noindent where $K_{EE}$ is the resistance at the E synapse. Therefore, combining Eqs. (\ref{eqn:1}-\ref{eqn:2}), the local membrane potential $V_{E}$ is expressed as
\begin{equation}
\label{eqn:3}
V_{E}=\frac{g_{E}K_{EE}\varepsilon_{E}}{1+g_{E}K_{EE}}.
\end{equation}

\noindent Similarly, the membrane potential measured at the soma $V_{S}$ in response to the synaptic input $g_{E}$ can be computed by
\begin{equation}
\label{eqn:4}
V_{S}=K_{ES}I_{E},
\end{equation}

\noindent where $K_{ES}$ is the transfer resistance between the E synapse and the soma. The combination of Eqs. (\ref{eqn:1}-\ref{eqn:4}) yields the somatic membrane potential in response to $g_{E}$ on the dendrite,
\begin{equation}
\label{eqn:5}
V_{S}=\frac{g_{E}K_{ES}\varepsilon_{E}}{1+g_{E}K_{EE}}.
\end{equation}

\noindent Now if we consider the \emph{soma} of the neuron to be electrically compact, and denote the current at the soma as $I^{eff}_{E}$, which is termed as the effective synaptic current (the same as $I^{(0)}_{E}$ in the main text) and is induced by the synaptic current $I_{E}$ (Eq. \ref{eqn:1}) from the dendrite, then the effective conductance, denoted by $g^{eff}_{E}$, satisfies
\begin{equation}
\label{eqn:6}
I^{eff}_{E}=g^{eff}_{E}(\varepsilon_{E}-V_{S}),
\end{equation}
where the superscript ``$eff$'' emphasizes the effective quantities at the soma. \\

\noindent Ohm's law links the somatic membrane potential with the somatic current as
\begin{equation}
\label{eqn:7}
V_{S}=K_{SS}I^{eff}_{E},
\end{equation}
where $K_{SS}$ is the resistance at the soma. \\

\noindent From Eqs. (\ref{eqn:6}-\ref{eqn:7}), we can obtain an expression for the effective E conductance $g^{eff}_{E}$ at the soma,
\begin{equation}
\label{eqn:8}
g^{eff}_{E}=\frac{V_{S}}{K_{SS}(\varepsilon_{E}-V_{S})}.
\end{equation}

\noindent Because $\varepsilon_{E}$ is relatively large compared with $V_{S}$, using Taylor expansion, Eq. (\ref{eqn:8}) can be approximated as
\begin{equation}
\label{eqn:9}
g^{eff}_{E} = \frac{V_{S}}{K_{SS}\varepsilon_{E}}\Big(1+\frac{V_{S}}{\varepsilon_{E}}+o\Big( \frac{V_{S}}{\varepsilon_{E}}\Big) \Big).
\end{equation}

\noindent Substituting Eq. (\ref{eqn:5}) into Eq. (\ref{eqn:9}), and assuming that $g_{E}$ is small, we have the effective E conductance at the soma,
\begin{equation}
\label{eqn:10}
g^{eff}_{E} = \frac{K_{ES}}{K_{SS}}g_{E}+o(g_{E}).
\end{equation}
where $o(g_{E})$ includes all high-order terms of the local conductance $g_{E}$.\\

\noindent Similarly, we can derive an expression for the effective I conductance at the soma,
\begin{equation}
\label{eqn:11}
g^{eff}_{I} = \frac{K_{IS}}{K_{SS}}g_{I}+o(g_{I}).
\end{equation}
where $o(g_{I})$ includes all high-order terms of the local conductance $g_{I}$. Therefore, to the first order accuracy of $g_{E}$ and $g_{I}$, the effective conductance can be viewed as a proportional indicator of the local postsynaptic conductance induced on the dendrite.

\subsection*{Determination of Effective Conductance}

We now demonstrate how to determine the effective conductances using voltage clamp when a neuron receives both E and I inputs on the dendrite. \\

\noindent By Ohm's law, we have the local E and I synaptic currents on the dendrite\\
\begin{equation}
\label{eqn:12}
I_{E}=g_{E}(\varepsilon_{E}-V_{E}),
\end{equation}
\begin{equation}
\label{eqn:13}
I_{I}=g_{I}(\varepsilon_{I}-V_{I}).
\end{equation}
Meanwhile, we can describe the local membrane potentials measured at the E synapse, the I synapse, and the soma as follows,
\begin{equation}
\label{eqn:14}
V_{E}=K_{EE}I_{E}+K_{IE}I_{I}+K_{SE}I_{inj},
\end{equation}
\begin{equation}
\label{eqn:15}
V_{I}=K_{EI}I_{E}+K_{II}I_{I}+K_{SI}I_{inj},
\end{equation}
\begin{equation}
\label{eqn:16}
V_{S}=K_{ES}I_{E}+K_{IS}I_{I}+K_{SS}I_{inj}.
\end{equation}
Here somatic voltage clamp is applied to hold the somatic membrane potential $V_{S}$ at a fixed level by injecting current $I_{inj}$ at the soma. The transfer resistance between locations A and B ($K_{AB}$) is the ratio of the voltage change in location B to the magnitude of the injected current in location A. These transfer resistances possess a reciprocal relation, i.e., the symmetric property, $K_{XY}=K_{YX}$, which is used in our calculation. Next, to obtain a relation between the somatic voltage and the synaptic current in the presence of the injected current, we solve Eqs. (\ref{eqn:12}-\ref{eqn:16}).\\

\noindent The somatic voltage can be obtained as
\begin{eqnarray}
\label{array:4}
V_{S}&=&\frac{g_{E}\varepsilon_{E}\Big(K_{ES}+g_{I}(K_{ES}K_{II}-K_{EI}K_{IS})\Big)+g_{I}\varepsilon_{I}\Big(K_{IS}+g_{E}(K_{IS}K_{EE}-K_{EI}K_{ES})\Big)}{1+g_{E}K_{EE}+g_{I}K_{II}+g_{E}g_{I}(K_{EE}K_{II}-K^2_{EI})} \nonumber \\
&+& \frac{I_{inj}\Big(K_{SS}+g_{E}(K_{EE}K_{SS}-K^2_{ES})+g_{I}(K_{II}K_{SS}-K^2_{IS}) \Big)}{1+g_{E}K_{EE}+g_{I}K_{II}+g_{E}g_{I}(K_{EE}K_{II}-K^2_{EI})} \\
&-&\frac{I_{inj}g_{E}g_{I}(K_{ES}^2K_{II}+K_{EE}K_{IS}^2+K_{EI}^2K_{SS}-K_{EE}K_{II}K_{SS}-2K_{EI}K_{ES}K_{IS} )}{1+g_{E}K_{EE}+g_{I}K_{II}+g_{E}g_{I}(K_{EE}K_{II}-K^2_{EI})}. \nonumber
\end{eqnarray}

\noindent On the other hand, under voltage clamp mode, the total current at the soma leads to the somatic voltage fixed at $V_{S}$,
\begin{equation}
K_{SS}(I^{inj}_{syn}+I_{inj})=V_{S}.  \nonumber \\
\end{equation}

\noindent Therefore, the synaptic current $I^{inj}_{syn}$ at the soma in the presence of the injected current can be obtained as
\begin{eqnarray}
\label{eqn:17}
I^{inj}_{syn}&=&\frac{V_{S}}{K_{SS}}-I_{inj}  \nonumber \\
&=&-\frac{g_{E}K^2_{ES}+g_{I}K^{2}_{IS}-g_{E}g_{I}(2K_{EI}K_{ES}K_{IS}-K^{2}_{ES}K_{II}-K^{2}_{IS}K_{EE})}{K_{SS}+g_{E}K_{EE}K_{SS}+g_{I}K_{II}K_{SS}+g_{E}g_{I}(K_{EE}K_{II}K_{SS}-K^2_{EI}K_{SS})} I_{inj}\\
&+&\frac{g_{E}\varepsilon_{E}K_{ES}+g_{I}\varepsilon_{I}K_{IS}+g_{E}g_{I}\Big(\varepsilon_{E}(K_{II}K_{ES}-K_{EI}K_{IS})+\varepsilon_{I}(K_{EE}K_{IS}-K_{EI}K_{ES})  \Big)}{K_{SS}+g_{E}K_{EE}K_{SS}+g_{I}K_{II}K_{SS}+g_{E}g_{I}(K_{EE}K_{II}K_{SS}-K^2_{EI}K_{SS})} \nonumber
\end{eqnarray}
where the superscript ``$inj$'' emphasizes that the synaptic current is measured in the presence of injected current given by the voltage clamp. We note that the synaptic current in Eq. (\ref{eqn:17}) depends on the injected current, indicating a nonlinear interaction between the synaptic currents from the dendrite and the injected current at the soma. We also note that both $V_{S}$ (in Eq. \ref{array:4}) and $I^{inj}_{syn}$ (in Eq. \ref{eqn:17}) are linear functions of $I_{inj}$, therefore, we can find a linear relationship between $V_{S}$ and $I^{inj}_{syn}$
\begin{equation}
\label{eqn:18}
I^{inj}_{syn}= -kV_{S}+b,
\end{equation}
which holds for arbitrary $I_{inj}$, and the corresponding holding potential $V_{S}$. \\

\noindent By comparing the coefficients of $I_{inj}$ on both sides of Eq. (\ref{eqn:18}), we can obtain the expression for the slope $k$ and the intercept $b$ as follows,
\begin{equation}
\!\!\!\!\!\!\!\!\!\!\!\!\!\!\!\!\!\!\!\!\!\!\!\!\!\!\!\!\!\!\!\!\!\!\!\!\!\!\!\!\!\!\!\!\!\!  k=\frac{g_{E}K^2_{ES}+g_{I}K^2_{IS}+g_{E}g_{I}(K_{ES}^2K_{II}+K^{2}_{IS}K_{EE}-2K_{EI}K_{ES}K_{IS})}{K_{SS}\Big(K_{SS}+g_{E}(K_{EE}K_{SS}-K^{2}_{ES})+g_{I}(K_{II}K_{SS}-K^2_{IS})+g_{E}g_{I}(K_{EE}K_{II}K_{SS}+2K_{EI}K_{ES}K_{IS}-K^2_{ES}K_{II}-K^2_{IS}K_{EE}-K^2_{EI}K_{SS})\Big)} \nonumber
\end{equation}

\begin{equation}
\!\!\!\!\!\!\!\!\!\!\!\!\!\!\!\!\!\!\!\!\!\!\!\!\!\!\!\!\!\!\!\!\!\!\!\!\!\!\!\!\!\!\!\!\!\! b=\frac{g_{E}\varepsilon_{E}K_{ES}+g_{I}\varepsilon_{I}K_{IS}+g_{E}g_{I}\Big(\varepsilon_{E}(K_{ES}K_{II}-K_{EI}K_{IS})+\varepsilon_{I}(K_{IS}K_{EE}-K_{EI}K_{ES})\Big)}{K_{SS}+g_{E}(K_{EE}K_{SS}-K^{2}_{ES})+g_{I}(K_{II}K_{SS}-K^2_{IS})+g_{E}g_{I}(K_{EE}K_{II}K_{SS}+2K_{EI}K_{ES}K_{IS}-K^2_{ES}K_{II}-K^2_{IS}K_{EE}-K^2_{EI}K_{SS})} \nonumber
\end{equation}

\noindent To the first order accuracy of $g_{E}$ and $g_{I}$, the expressions of $k$ and $b$ reduce to
\begin{eqnarray}
\label{eqn:kkk}
k&=&\Big(\frac{K_{ES}}{K_{SS}}\Big)^2g_{E}+\Big(\frac{K_{IS}}{K_{SS}}\Big)^2g_{I} +o(g) \nonumber\\
&=&\Big(\frac{K_{ES}}{K_{SS}}\Big)g^{eff}_{E}+\Big(\frac{K_{IS}}{K_{SS}}\Big)g^{eff}_{I} +o(g)
\end{eqnarray}

\begin{eqnarray}
\label{eqn:bbb}
b&=&\Big(\frac{K_{ES}}{K_{SS}}\Big)g_{E}\varepsilon_{E}+\Big(\frac{K_{IS}}{K_{SS}}\Big)g_{I}\varepsilon_{I} +o(g)\nonumber \\
&=& g^{eff}_{E}\varepsilon_{E}+g^{eff}_{I}\varepsilon_{I} +o(g)
\end{eqnarray}

\noindent The above results show that \textit{the slope of the I-V relation is not equal to the total effective conductance but the intercept equals the reversal current}. Therefore, the I-V relation allows only the value of the intercept $b$ to measure the E and I conductances but not the value of the slope $k$ which contain generally unknown factors $K_{ES}/K_{SS}$ and $K_{IS}/K_{SS}$. Because we have two unknown variables $g^{eff}_{E}$ and $g^{eff}_{I}$ here, in order to solve them, we need to change the reversal potential to provide at least two equations of I-V relations.\\

\subsection*{Error Estimation}
From the above analysis, the intercept method is of a first-order accuracy of the local conductance $g_{E}$ and $g_{I}$ in the effective conductance measurement. In addition, Eqs. (\ref{eqn:kkk}-\ref{eqn:bbb}) indicate that, to the first order approximation, the E and I conductances are
\begin{equation}
g_{E}^{eff}=\frac{b(\frac{K_{IS}}{K_{SS}})+k\varepsilon_{I}}{(\frac{K_{ES}}{K_{SS}})\varepsilon_{I}-(\frac{K_{IS}}{K_{SS}})\varepsilon_{E}},
\nonumber
\end{equation}
\begin{equation}
g_{I}^{eff}=\frac{b(\frac{K_{ES}}{K_{SS}})+k\varepsilon_{E}}{(\frac{K_{IS}}{K_{SS}})\varepsilon_{E}-(\frac{K_{ES}}{K_{SS}})\varepsilon_{I}}.
\nonumber
\end{equation}
Considering a special case when the E and I inputs are elicited at the same location, then we have $K_{ES}=K_{IS}=\alpha K_{SS}$, where $\alpha$ is a number less than one, which reflects the effect of spatial dependence of transfer resistance. The above results can be simplified as
\begin{equation}
\label{eqn:ge_exp}
g_{E}^{eff}=\frac{b+k\varepsilon_{I}/\alpha}{\varepsilon_{I}-\varepsilon_{E}},
\end{equation}
\begin{equation}
\label{eqn:gi_exp}
g_{I}^{eff}=\frac{b+k\varepsilon_{E}/\alpha}{\varepsilon_{E}-\varepsilon_{I}}.
\end{equation}
If we use the traditional slope-and-intercept method, the estimated conductance is obtained by assuming $\alpha=1$ always in Eqs. (\ref{eqn:ge_exp}-\ref{eqn:gi_exp}). Therefore, the absolute errors for the conductance measurements between the two methods are
\begin{equation}
\label{eqn:smaller1}
\Delta g_{E}^{eff}= \frac{k\varepsilon_{I}(1-\frac{1}{\alpha})}{\varepsilon_{I}-\varepsilon_{E}},
\end{equation}
\begin{equation}
\label{eqn:smaller2}
\Delta g_{I}^{eff}= \frac{k\varepsilon_{E}(1-\frac{1}{\alpha})}{\varepsilon_{E}-\varepsilon_{I}}.
\end{equation}
Clearly, the ratio between the two absolute errors is $\Delta g_{E}^{eff}:\Delta g_{I}^{eff}=-\varepsilon_{I}:\varepsilon_{E}$. From  Eqs. (\ref{eqn:smaller1})-(\ref{eqn:smaller2}), $\Delta g_{E}^{eff}$ and $\Delta g_{I}^{eff}$ are both negative, which indicates that the conductance measured by SIM is always smaller than the true effective conductance. This is consistent with our experimental and simulation results in the main text.

\subsection*{Relation between local conductance and conductance determined by SIM}
Here we use the method of proof by contradiction to demonstrate that there does not exist a transform between the local E (I) conductance and the E (I) conductance determined by SIM. \\

\noindent Let's assume there are such transforms that link the conductances determined by SIM and the local ones, i.e., $g^{SIM}_{E}=F(g_{E})$, $g^{SIM}_{I}=G(g_{I})$. On the one hand, according to SIM, we have
\begin{equation}
k=F(g_{E})+G(g_{I}), \nonumber
\end{equation}
\begin{equation}
b=F(g_{E})\varepsilon_{E}+G(g_{I})\varepsilon_{I}. \nonumber
\end{equation}
By expanding $F(g_{E})$ and $G(g_{I})$ as Taylor series to the first order accuracy, we have
\begin{equation}
\label{slope1}
k=F_0+G_0+F_1 g_{E}+G_1 g_{I}+o(g),
\end{equation}
\begin{equation}
\label{intercept1}
b=F_0\varepsilon_{E}+G_0\varepsilon_{I}+F_1g_{E}\varepsilon_{E}+G_1g_{I}\varepsilon_{I}+o(g).
\end{equation}
where $F_0$, $F_1$, $G_0$, and $G_1$ are the coefficients of the Taylor series.
On the other hand, from Eqs. (\ref{eqn:kkk})-(\ref{eqn:bbb}), we have $g_{E}$ and $g_{I}$ satisfying
\begin{equation}
\label{slope2}
k=\Big(\frac{K_{ES}}{K_{SS}}\Big)^2g_{E}+\Big(\frac{K_{IS}}{K_{SS}}\Big)^2g_{I} +o(g) \\
\end{equation}
\begin{equation}
\label{intercept2}
b=\Big(\frac{K_{ES}}{K_{SS}}\Big)g_{E}\varepsilon_{E}+\Big(\frac{K_{IS}}{K_{SS}}\Big)g_{I}\varepsilon_{I} +o(g) \\
\end{equation}
By comparing the coefficients in Eq. (\ref{slope1}) and those in Eq. (\ref{slope2}), we have
\begin{equation}
\label{k_co}
F_1=\Big(\frac{K_{ES}}{K_{SS}}\Big)^2, \ \ \ \ G_1=\Big(\frac{K_{IS}}{K_{SS}}\Big)^2.
\end{equation}
By comparing the coefficients in Eq. (\ref{intercept1}) and those in Eq. (\ref{intercept2}), we have
\begin{equation}
\label{b_co}
F_1=\frac{K_{ES}}{K_{SS}}, \ \ \ \ G_1=\frac{K_{IS}}{K_{SS}}.
\end{equation}
Eq. (\ref{k_co}) contradicts to Eq. (\ref{b_co}), which proves that there does not exist such transforms between the local conductance and the conductance determined by SIM. Therefore, the conductance determined by SIM is incapable of reflecting the input conductance information, thus has no clear biological interpretation.

\end{document}